\documentclass{aastex62}

\usepackage{graphicx}
\usepackage{xcolor}

\definecolor{sof}{rgb}{0.4, 0.0, 0.6}

\shorttitle{9 Axes of Merit}
\shortauthors{Sheikh}

\begin{document}

\title{Nine Axes of Merit for Technosignature Searches}
\author[0000-0001-7057-4999]{Sofia Z. Sheikh}
\address{Department of Astronomy \& Astrophysics and \\ Center for Exoplanets and Habitable Worlds\\ 525 Davey
Laboratory, The Pennsylvania State University, University Park, PA, 16802, USA}

\begin{abstract}

The diverse methodologies and myriad orthogonal proposals for the best technosignatures to search for in SETI can make it difficult to develop an effective and balanced search strategy, especially from a funding perspective. Here I propose a framework to compare the relative advantages and disadvantages of various proposed technosignatures based on nine ``axes of merit''. This framework was first developed at the NASA Technosignatures Workshop in Houston in 2018 and published in that report. I give the definition and rationale behind the nine axes as well as the history of each axis in the SETI and technosignature literature. These axes are then applied to three example classes of technosignature searches as an illustration of their use. An open-source software tool is available to allow technosignature researchers to make their own version of the figure.
\end{abstract}


\section{Introduction}

Proposed searches for technosignatures range from radio wavelengths to gamma rays, take advantage of almost every astronomical dataset, and use interdisciplinary methodologies in such a way that comparing the merits of two dissimilar searches, even if they're ostensibly in the same field, can be an extremely difficult task. Each SETI practitioner has a different answer for the best strategy to find ETI, often in her own wavelength. Much of the SETI literature engages in promoting the values of a particular search strategy. While ``figures of merit'' have been proposed to compare radio SETI searches (see, e.g., \citet{enriquez2017breakthrough}, \citet{Wright2018a}), comparing searches across disparate modalities is a more difficult and less frequently attempted task\footnote{Works in this spirit include \citet{arnold_2013} and the series of papers beginning with \citet{Hippke2017}}. 

A productive discussion from the NASA Technosignatures Workshop in Houston, Texas in late 2018 sparked the idea of comprehensively comparing searches in a more inclusive technosignature framework. Inspired by individual metrics proposed by the workshop presenters, I created the "Axes of Merit" which were then incorporated into the introduction of the final workshop report \citep{participants2018nasa}. Since then, the axes have appeared in other venues as well \citep{Berdyugina2019AbSciCon, Angerhausen2019AbSciCon}. In response to the apparent usefulness of the idea, I here formalize the axes and provide a more in-depth description than that provided in the Houston Report, give credit and historical context to the intellectual lineage of these ideas, demonstrate how this framework can be applied to technosignature searches with concrete examples, and provide a tool to create publication-quality illustrations of this concept with an updated graphic.

The Axes of Merit themselves are described in Section \ref{sec: axes_description}, some examples of their use are given in Section \ref{sec: qual_examples}, a discussion of this framework, including insights, caveats, and limitations, is provided in Section \ref{sec: discussion}, and information about an open-source figure-generation tool is provided in Section \ref{sec: figgen}.

\section{The Nine Axes of Merit}
\label{sec: axes_description}

The nine axes of merit for technosignature searches are as follows:

\begin{enumerate}
    \item Observational Capability
    \item Cost
    \item Ancillary Benefits
    \item Detectability
    \item Duration
    \item Ambiguity
    \item Extrapolation
    \item Inevitability
    \item Information
\end{enumerate}

\begin{figure}
    \centering
    \includegraphics[width=0.77\textwidth]{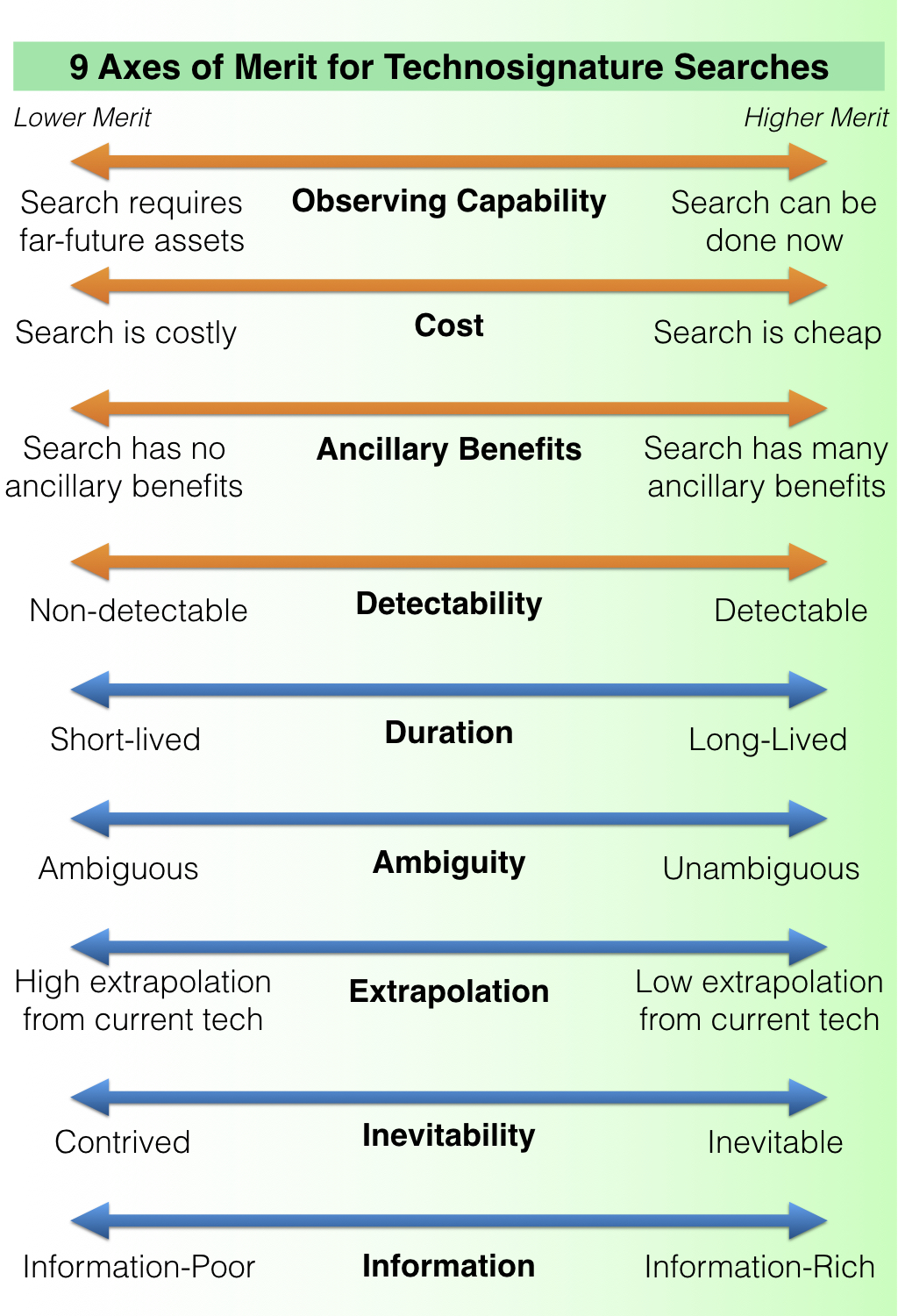}
    \caption{A visual representation of the Nine Axes of Merit described in Section \ref{sec: axes_description}.}
    \label{fig:axes_of_merit}
\end{figure}

\subsection{Observing Capability}
Observing capability refers to the technological ability of astronomy as a whole at the time a search for the technosignature is proposed. This is often driven by the difficulty of developing and deploying a new technology to perform an efficient and thorough search. \citet{Klein1991} noted that a successful SETI search requires a match between the technology of the transmitter and the technology of the receiver; in this framework, we have only been technologically able to perform such searches for 75 years. The argument that SETI strategy should be dictated by our current capabilities is also made by \citet{Stull1979}.

\subsection{Cost}
Cost must be considered in any astronomical program, but given SETI's history (and present reality) of uncertain funding it has traditionally been especially prioritized by technosignature researchers. Cost, in this context, includes not only financial costs, but also telescope time, computing time, and other opportunity costs. SETI has often been forced to prioritize this axis highly at the expense of other axes, for example using ``parasitic'', ``commensal'', or ``piggybacking'' strategies (e.g. \citet{Bowyer1983}) to allow SETI programs to be performed concurrently with other, non-SETI research (which leave the programs at the mercy of the scientific choices made for the non-SETI research). The funding problem for SETI is severe enough that it has been considered in more quantitative terms as well \citep{Lingam2018}.

The idea that cost should be a strong driver in technosignature research has been identified by \citet{davies2013searching}, which makes the argument that a search with a low cost should be prioritized even if the signature in question appears to have low plausibility at the outset. Cost has even been suggested as an important factor in the motivation of the \textit{transmitter} (e.g. \citet{Benford2008}). 

\subsection{Ancillary Benefits}
Many technosignature searches involve surveys that can be used for other purposes, or can be expected to discover anomalies of significant and potentially transformative astrophysical importance. We should prioritize searches that satisfy Freeman Dyson's First Law of SETI Investigations: ``Every search for alien civilizations should be planned to give interesting results even when no aliens are discovered'' \citep{participants2018nasa}. The ancillary benefits of a search might be technological, as in the wide-band data recording hardware for the Green Bank telescope developed by Breakthrough Listen which has enabled significant discovery \citep{Macmahon2018}. Finally, the ancillary benefits could appear in non-STEM contexts such as philosophy, education, and policy (e.g. \citet{Tough1998}).

\subsection{Detectability}
As with biosignatures, a useful technosignature is one that produces a strong signal relative to the background noise. For instance, the spillover energy from a directed energy drive for interstellar spacecraft would be extremely bright and easily detected \citep{Harris1986}, whereas the transit signature of artificial satellites might be extremely subtle \citep{socas2018possible}.

\subsection{Duration}
Duration refers to the length of time that a technosignature would be detectable; this idea is central to the transmitter lifetime $L$ in the Drake Equation \citep{Drake1965}. For periodic signatures, the duty cycle also contributes to the duration axis \citep{Wright2018a}. For instance, signs of propulsion of interstellar craft might occur only in bursts, and so require long and repeated observations of the same location before being discovered. Some technosignatures may only occur for a brief period in a technological species' development, and so only be present among a small fraction of host stars (for example, strong detectable effects on a planet's atmosphere might be a predictable and temporary planetary transition \citep{Frank2017}). Persistent technosignatures, however (e.g. continually-transmitting beacons or waste heat) may require only a single observation. Strong performance on the duration axis is also a reason to potentially prefer technosignatures to biosigantures \citep{Cirkovic2019}.

\subsection{Ambiguity}
As with biosignatures such as atmospheric oxygen, which has both biotic and abiotic sources \citep{Meadows2018}, some technosignatures might be easily mistaken for natural phenomena unrelated to life. For instance, waste heat from technology has a similar observational signature to astrophysical dust, which makes it difficult to differentiate them \citep{Dyson1966, Wright2014}. Extremely narrow-band radio emissions, on the other hand, do not have a natural source, and thus provide a thoroughly unambiguous signature of technology \citep{COCCONI1959}. 

\subsection{Extrapolation}
We can be more confident that a technosignature might exist if we already understand and use the underlying technology on Earth. Laser and radio signals are popular targets of searches for technosignatures because humanity would be capable of detecting its own such signals at interstellar distances \citep{drake1973interstellar}. Towards the other end of the axis, the creation of a Dyson sphere would greatly exceed humanity's current capabilities. Some proposed technosignatures (such as exotic forms of propulsion) not only require extrapolation beyond our current capabilities, but also beyond our current understanding of fundamental physics. \citet{socas2018possible} emphasizes looking for ``technomarkers'' that could be produced with our own technological abilities. Other authors have highlighted the potential dangers of extrapolating forward on Myr or greater timescales when we only have on the order of $10^{-2}$ Myr of recorded human history and technological development on Earth \citep{Mix2019AbSciCon}.  

\subsection{Inevitability}
Given a distribution of technological ETIs, what fraction of them will create a given technosignature? Technosignatures which appear in all of them should be highly prioritized, while technosignatures whose creation relies on assumptions about the behaviour, sociology, or psychology of an ETI should be penalized. This has been argued many times \citep{Kuiper1977, Stull1979}. An ``agnostic'' technosignature search (e.g. signal-shape agnostic communication searches) will score better along this axis. Waste heat is an inevitable consequence of energy use according to fundamental physics, making for a particularly robust technosignature \citep{Dyson1960, Wright2014}. Conversely, the decision to send an intentional transmission relies on assumptions of knowledge and motivation of an ETI for which we have no way to determine a quantitative model.

\subsection{Information}
Though many SETI practitioners focus their professional energy towards making the first detection, scientifically, the value of the discovery of a technosignature would be proportional to the amount of information that could be derived from it. An information-rich technosignature, such as the discovery of an extraterrestrial artifact within the solar system would enable in-situ analysis, leading to scientific gains far surpassing what we could hope to learn from most other technosignatures. A ``bare contact'' signature \citep{Cirkovic2018}, on the other hand, would lead to a binary yes or no answer to the question of the existence of technological life, but would not provide as much additional scientific insight. 

\section{Qualitative Axis Values for Three Selected Technosignature Search Strategies}
\label{sec: qual_examples}

The placement of a given search on the Axes of Merit is necessarily a subjective exercise; the three examples listed below are qualitative.

\begin{figure}[ht]
    \centering
    \includegraphics[height = 0.475\textheight]{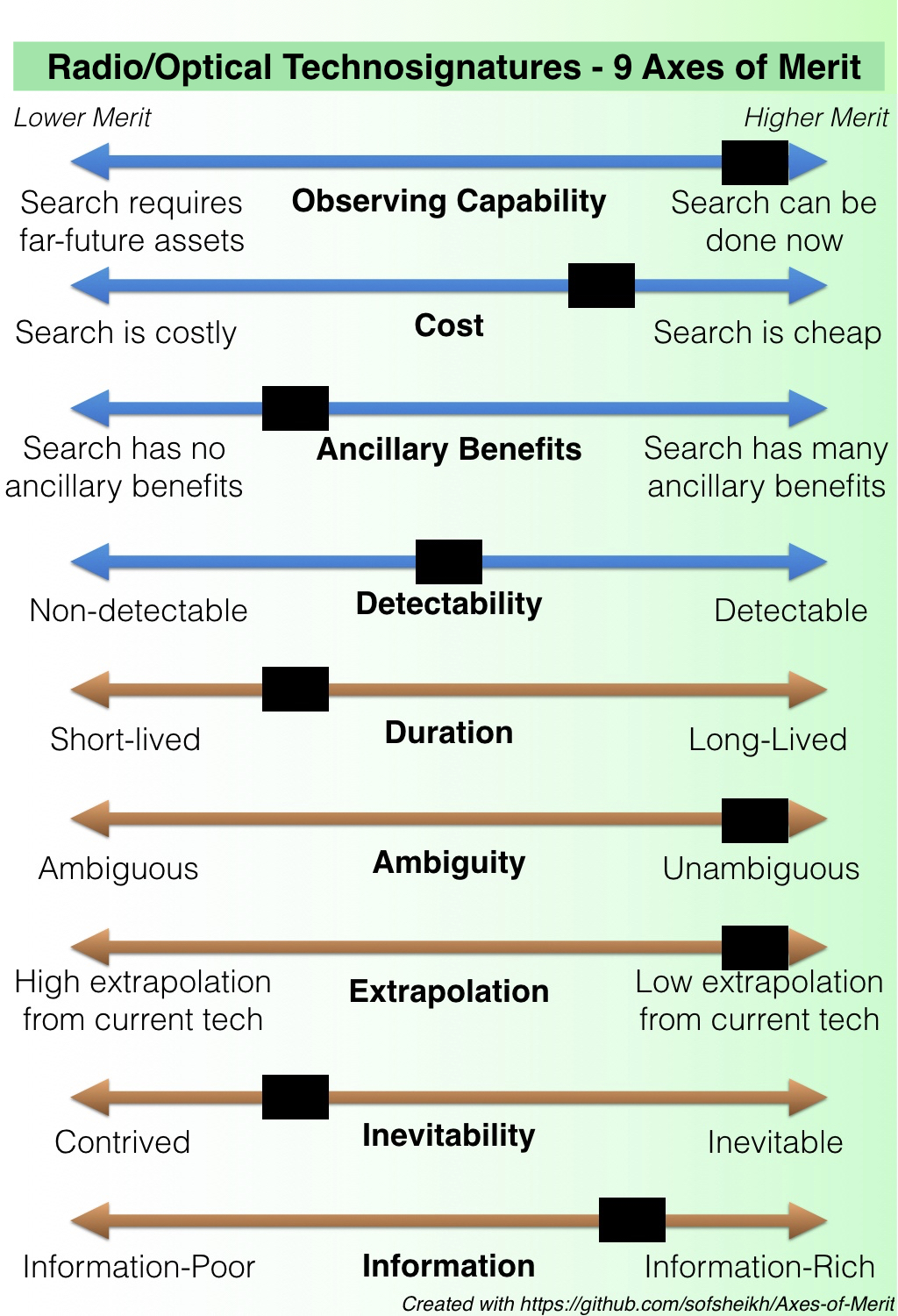}
    \caption{Radio and optical communication}
    \label{subfig:radio_optical}
\end{figure}

\subsection{Radio and Optical Communication}
\label{subsec: radiooptical}

Since the foundation of the discipline in the 1960s, the majority of efforts in traditional SETI have focused on searches for intentional radio and optical communication \citep{COCCONI1959, Drake1961, SCHWARTZ1961}. These efforts originated with the development of powerful electromagnetic transmission/reception technology. Both radio and optical approaches share the same philosophy: search for technologically-generated electromagnetic signals that are compressed in either time or frequency beyond what would be expected from a natural astrophysical source. The transmitter could be directing the radiation in a beam which contains the Earth or could be emitting isotropically.

Radio and optical communication searches are qualitatively ranked on the Nine Axes in Figure \ref{subfig:radio_optical}. I explain the rationale behind the rankings below.
\begin{enumerate}
    \item These searches have often been favored because of their strong performance in Observing Capability.
    \item Cost for small projects is relatively low, consisting of getting telescope time, commensal search permission, or access to archival data from existing instrumentation. However, the most comprehensive searches in each wavelength \citep{Macmahon2018, wright2018panoramic} have required the development of e.g. new backends and receivers, increasing the cost. 
    \item These searches specifically look in areas of parameter space that are not occupied by known astrophysical phenomena (causing them to score perfectly on the Ambiguity axis), but limit the amount of Ancillary Benefits that can come out of these projects. Some of these parameter spaces, however, are not as empty as originally thought; the discovery of fast radio bursts \citep{Lorimer2007} illustrates the potential Ancillary Benefits of such searches. 
    \item The Detectability of intentional electromagnetic signals is moderate; the signals are meant to be detected, so they should be relatively strong and undisguised but this also depends strongly on distance, transmitter size, and frequency of transmitters in the galaxy/universe.
    \item Radio and optical communication require a constant power source, which places limits on the Duration. A transmitter could outlive its host ETI, but transmitters that need active maintenance would track the lifetime of the host ETI, potentially leading to a short median duration.
    \item These searches excel in being unambiguous as is discussed with Ancillary Science above. 
    \item Extrapolation from current Earth technologies can be as low as zero for some targets, and this has been the case since the 1960s. Previous searches have often placed upper limits on transmitter power a few times above current Earth capabilities. Searches for extragalactic transmitters score much more poorly on this axis because of their Kardashev II-III levels of required energy \citep{kardashev1964}, however such searches are much less common and 
    I do not consider them the ``default'' for radio and optical communication.
    \item A common criticism of radio and optical communication is that they are strongly dependent on our ideas of technological development based on Earth history. It has been argued that electromagnetic radiation may not be a long, efficient, or necessary phase in the technological development of another ETI, and thus not Inevitable \citep{Forgan2011}. In addition, looking for intentional communications requires motivation to initiate a transmission, which assumes some sort of sociological reason to do so.
    \item Unlike most other SETI strategies, which tend to focus on unintentional technosignatures, radio and optical communication assume intentionality. This allows the potential Information content of the signal to be very large (as in the case of a decodable transmission), with some additional time component. Alternately, ``beacons'' \citep{Wright2018} would carry no information content.
\end{enumerate}

\subsection{Waste Heat from Megastructures}
\label{subsec: wasteheat}

\begin{figure}[ht]
    \centering
    \includegraphics[height=0.475\textheight]{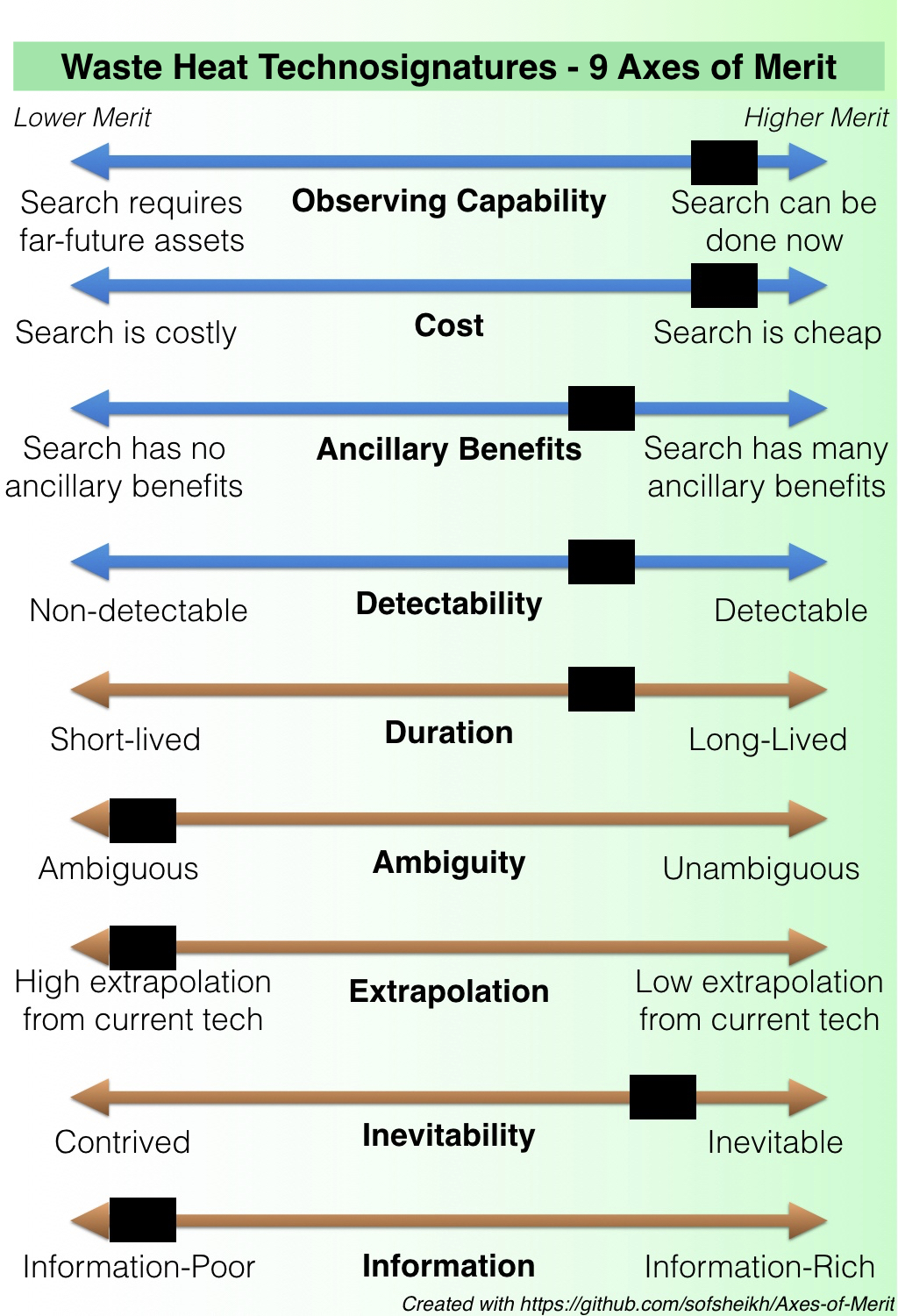}
    \caption{Waste heat}
    \label{subfig:waste_heat}
\end{figure}

\citet{Dyson1960} first introduced the idea of searching for megastructures. Here, I define megastructures as technological artifacts outside the solar system that are large enough to be remotely detectable due to their size alone, built by ETI for any purpose. It has been theorized that megastructures could be constructed for energy generation or as artificial habitable environments \citep{Dyson1960, Niven1970}, or could be used for communication \citep{Arnold2005} or computation \citep{bradbury2001matrioshka}. These megastructures could be observed in infrared wavelengths by the waste heat that they emit or in visible wavelengths by the starlight that they block in transit; here, I will focus on waste heat. Searches for megastructures have been rarely performed in the literature \citep{Jugaku1997, Carrigan2009, Wright2014} but often discussed in theory.

Waste-heat searches are qualitatively ranked on the Nine Axes in Figure \ref{subfig:waste_heat}. I explain the rationale behind the rankings below.
\begin{enumerate}
    \item There are many existing datasets that have not been searched for megastructure waste heat. The issue is not a lack of data or resources but instead a paucity of funded studies on the topic. No new instrumentation is needed to perform this work; we have the ability to detect megastructures with current Observing Capabilities.
    \item The Cost of performing archival data searches is extremely low.
    \item There are ample Ancilliary Benefits of a search for waste heat, as it will detect any object with an infrared excess. This benefits stellar astrophysics, planet formation, and studies of the interstellar medium.
    \item Waste heat signatures are quite Detectable, with potentially strong signal-to-noise ratios to differentiate from the null hypothesis.
    \item Given the extent of deep time and the unconstrained lifetimes of ETIs, it has been proposed that a sort of ``archaeology'' will be needed to understand the first SETI detections, as they are likely to be from extinct ETIs \citep{Carrigan2010}. Megastructure waste heat, as a product of a physical artifact, would greatly outlast the lifetime of its ETI creators and thus score well in Duration.
    \item Mining datasets for objects with infrared excess reveals dusty regions, protoplanetary disks, and other objects of astrophysical interest in addition to megastructure candidates. Other observational methods and further modelling are required to break the substantial Ambiguity between candidates and these other astrophysical objects.
    \item Humanity has never created a megastructure as defined in this section. While objects large enough to make an effect on a transit detection are more achievable than megastructures detectable via waste heat, both are extremely far-future technologies and would require a large degree of Extrapolation.
    \item The Inevitability of the waste heat technosignature has been touted as its most important feature. With only a single assumption, that the laws of thermodynamics are always valid, a technosignature \textit{must} be produced if a megastructure is created. However, this score also must reflect the likelihood of the existence of the megastructure in the first place, damping the usefulness of the signature from an Inevitability perspective.
    \item Megastructure waste heat is an extremely Information-Poor technosignature. A distant detection will result in little more than the knowledge that an artificial artifact exists (and whatever information can be observationally derived about its properties), with no obvious avenue for \textit{in situ} follow-up.
\end{enumerate}

\subsection{Solar System Artifacts}
\label{subsec: artifacts}

\begin{figure}[ht]
    \centering
    \includegraphics[height=0.475\textheight]{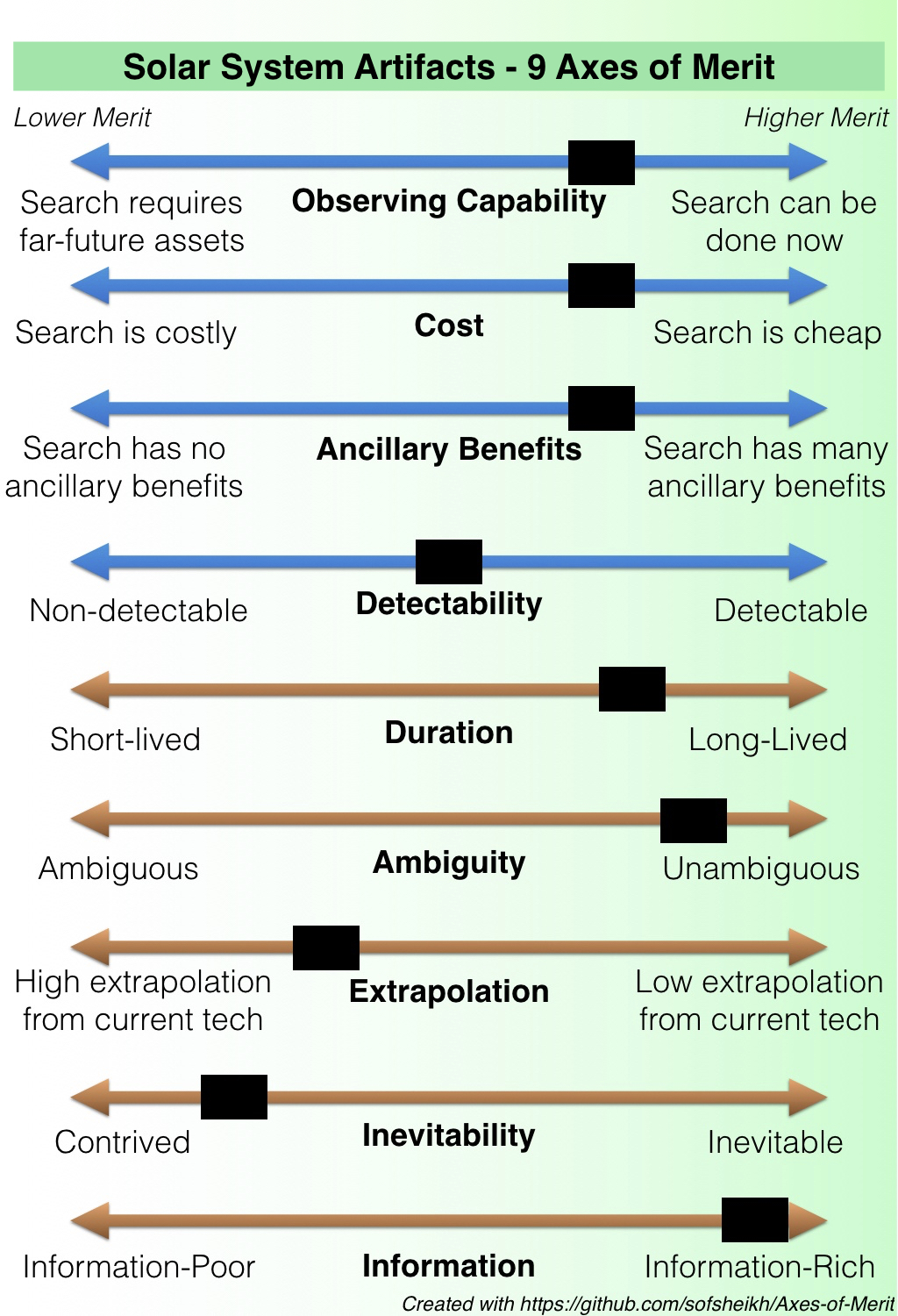}
    \caption{Solar system artifacts}
    \label{subfig:artifacts}
\end{figure}

For the purposes of this section, a solar system artifact is a technologically created non-human object, substance, pattern or process that exists within the boundaries of the solar system \citep{Bracewell1960}. This includes not just physical artifacts, but also technological ichnofossils (the indirect impact of technology on the geological record observed through trace fossils) and geochemical signals. Many possibilities for the motivation behind the creation of these artifacts have been advanced in the literature, including exploration (like an ETIs Breakthrough Starshot \citep{daukantas2017breakthrough}) and contact \citep{freitas1980interstellar}. This class of technosignatures includes artifacts from both non-solar ETIs and prior indigenous technological species within the solar system \citep{wright2018prior}, and artifacts that are free-floating and surface-dwelling.

Solar system artifact searches are qualitatively ranked on the Nine Axes in Figure \ref{subfig:artifacts}. I explain the rationale behind the rankings below.
\begin{enumerate}
    \item Our Observing Capabilities within the solar system are already sufficient to perform many solar system artifact searches that have not yet been done, and our current search completion is extremely low \citep{Haqq-Misra2012}.
    \item The Cost of searching for solar system artifacts is relatively low because, in many cases, it relies on existing instrumentation and resources (e.g. \citet{davies2013searching}).
    \item The Ancillary Benefits of an artifact search can be substantial due to the synergies with existing missions. Searching for artificial features on terrestrial bodies could result in new knowledge about planetary surface processes. Similarly, free-floating artifact searches work nicely in tandem with small body research.
    \item The Detectability of solar system artifacts varies widely, so it is difficult to place it on the axis. The relatively small distances involved enhance the detectability, but the sizes and ages of potential technosignatures could negate this benefit.
    \item As explained in Section \ref{subsec: wasteheat}, physical artifacts will likely be longer lived than their electromagnetic counterparts. Thus, solar system artifacts score well along the Duration axis. 
    \item Solar System Artifacts are strongly unambiguous technosignatures because we will be able to further analyze and vet potential candidates with both remote and \textit{in situ} techniques in a way that cannot be done with searches outside the solar system. Perfectly unambiguous signatures, however, may not be possible due to limited data and the slow destruction of the artifact with time.
    \item The furthest human-made artifact in the universe is Voyager 1, which has only recently passed beyond the heliopause\footnote{https://voyager.jpl.nasa.gov/mission/status/}. No human technology has ever travelled between stellar systems; solar system artifacts score rather poorly on Extrapolation.
    \item Some authors have made claims about the universality of probes as an inevitable technosignature \citep{Bracewell1960}, and the lack of their observation a sure sign of the rarity of intelligent life \citep{Hart1975}. However, there is no physically-motivated reason for the construction of interstellar artifacts to be a common phenomenon, and thus they are not an Inevitability.
    \item Solar system artifacts are Information-Rich technosignatures due to our ability to examine them \textit{in situ}. While engineered probes would have different information content than e.g. the discovery of an ancient mining site on an asteroid, both could be closely examined with and informed by the methods and tools used in archaeology on Earth \citep{McGee2010, denninghumans} to extract information that would be impossible to discern for any other class of technosignature.
\end{enumerate}

\section{Discussion}
\label{sec: discussion}
\subsection{Insights from the Axis Framework}
\label{subsec:consequences}

Historically, an extremely biased weighting of the axes has been used as a model for technosignature searches; Axes 1 and 2 (and to some degree 3), dealing with the practical aspects of any technosignature search, have always been disproportionately weighted because of the paucity of funding and support in the field. Thus there exists a large, unperformed set of searches that would score better along the science-focused axes than most studies to date. This makes the field ripe with ``low-hanging fruit'' in a way most mature astrophysical disciplines no longer are.

At the same time, there are still many searches that prioritize the extreme positive end of those first three axes which have never been performed: most notably, searches through archival data from surveys for other astronomical phenomena. These datasets could contain technosignatures as proposed in Section \ref{sec: qual_examples}, or techosignatures that appear as ``nature-plus'' --- astrophysical objects, studied in existing bands with existing instrumentation, that do something that is physically impossible without technological intervention \citep{Davies2010}.

The data already exists and only requires robust upper limits to be calculated in the context of a technosignature search. Studies of this kind would likely employ machine learning methods, especially anomaly detection, and image processing methods.

\subsection{Limitations of and Caveats to the Axis Framework}
\label{subsec: caveats}

No conceptual framework is without limitation --- some of the larger ones are addressed here.

These axes are not entirely independent of each other. Contrived proposals for technosignatures with high extrapolation from current Earth capabilities can be almost arbitrarily detectable --- proposals for e.g. extremely large-scale astroengineering projects make this trade-off. 

With such a wide variety of potential searches and no priors, the axes should not be used to entirely exclude areas of search. While prioritizing searches is important, especially with limited resources, breadth will serve us better than depth in technosignature searches.

The axes will not capture all of the considerations in a search, but they capture the most fundamental ones. Other suggestions for axes were rejected from the final model as being too specific. One of these considerations was Potential for Concealment: if an ETI did not want to reveal its presence, how much could it prevent or hide the technosignature? This requires an assumption of concealment as a fundamental driving motivation and is closely linked to Inevitability. Another consideration was Physical Volume: should searches that cover more physical space be prioritized? The inclusion of this axis requires an inherent assumption of the rarity of ETI. A final suggestion was Size of Search Space: how difficult would it be to reach some degree of completion with the proposed technosignature? This proposed axis is strongly degenerate with Observing Capability and Detectability. For these reasons, none of these additional axes were included in the final model.

The Nine Axes, in their current form, cannot be used as a quantitative measure of the ``quality'' of a technosignature search. Even if scores were to be assigned, the weights given to each axis will depend on one's priors. These priors include ideas about the occurrence rate of ETI in the universe, the longevity of ETIs themselves, and the level of energy consumption and redirection of which the ETI is capable. Nothing is known about any of these variables, so these priors will be necessarily personal, philosophical, and ultimately subjective. Instead the axes should be used as an illustrative tool to motivate the choice of a particular search strategy and openly communicate its shortcomings.

Finally, a similar framework might be used to rank potential biosignatures. This is a potentially fruitful topic for future study but beyond the scope of this work.

\section{Figure Generation Tool}
\label{sec: figgen}

To make it easier to apply and use the framework of the Nine Axes of Merit, I have written a tool to produce versions of the plots in this paper with customizable axis values. The software tool is open-source and publicly-accessible\footnote{\url{https://github.com/sofsheikh/Axes-of-Merit}} and the plots that it creates can be used to standardize and compare technosignature proposals, presentations, and papers.

\section{Acknowledgements}

I would like to thank the organizers of and participants in the NASA Technosignatures Workshop 2018, at which this concept was invented. I would specifically like to acknowledge Adam Frank for his inspiring input on the Detectability and Inevitability axes and his gracious peer-review, David Kipping and Sara Walker for discussions that inspired the Ambiguity Axis, and Jason Wright for references and feedback. Finally, I would like to acknowledge the Breakthrough Listen Foundation and the SETI Institute for providing me a workspace and other support during Summer 2019.

\bibliographystyle{aasjournal}
\bibliography{references}

\begin{thebibliography}{}
\expandafter\ifx\csname natexlab\endcsname\relax\def\natexlab#1{#1}\fi
\providecommand{\url}[1]{\href{#1}{#1}}

\bibitem[{Angerhausen(2019)}]{Angerhausen2019AbSciCon}
Angerhausen, D. 2019, abSciCon Seattle

\bibitem[{Arnold(2013)}]{arnold_2013}
Arnold, L. 2013, International Journal of Astrobiology, 12, 212

\bibitem[{Arnold(2005)}]{Arnold2005}
Arnold, L. F.~A. 2005, The Astrophysical Journal, 627, 534.
\newblock \url{http://stacks.iop.org/0004-637X/627/i=1/a=534}

\bibitem[{Benford {et~al.}(2008)Benford, Benford, \& Benford}]{Benford2008}
Benford, G., Benford, J., \& Benford, D. 2008, arXiv pre-print, 18.
\newblock \url{http://inspirehep.net/record/800296/
  http://arxiv.org/abs/0810.3966{\%}5Cnhttp://arxiv.org/pdf/0810.3966}

\bibitem[{Berdyugina(2019)}]{Berdyugina2019AbSciCon}
Berdyugina, S. 2019, abSciCon Seattle

\bibitem[{Bowyer {et~al.}(1983)Bowyer, Zeitlin, Tarter, Lampton, \&
  Welch}]{Bowyer1983}
Bowyer, S., Zeitlin, G., Tarter, J., Lampton, M., \& Welch, W.~J. 1983, Icarus,
  53, 147.
\newblock
  \url{https://www.sciencedirect.com/science/article/abs/pii/0019103583900283}

\bibitem[{Bracewell(1960)}]{Bracewell1960}
Bracewell, R.~N. 1960, Nature, 186, 670.
\newblock \url{http://www.nature.com/articles/186670a0}

\bibitem[{Bradbury {et~al.}(2001)}]{bradbury2001matrioshka}
Bradbury, R.~J., {et~al.} 2001, unpublished

\bibitem[{Carrigan(2009)}]{Carrigan2009}
Carrigan, R.~A. 2009, Astrophysical Journal, 698, 2075.
\newblock \url{http://arxiv.org/abs/0811.2376
  http://dx.doi.org/10.1088/0004-637X/698/2/2075
  http://stacks.iop.org/0004-637X/698/i=2/a=2075?key=crossref.57dc0d2b5e75d1e21b10269a7716dd0e}

\bibitem[{Carrigan(2010)}]{Carrigan2010}
---. 2010, JBIS - Journal of the British Interplanetary Society, 63, 90.
\newblock \url{http://arxiv.org/abs/1001.5455}

\bibitem[{{\'{C}}irkovi{\'{c}}(2018)}]{Cirkovic2018}
{\'{C}}irkovi{\'{c}}, M. 2018, {The Great Silence: Science and Philosophy of
  Fermi's Paradox} (Oxford University Press).
\newblock
  \url{https://books.google.com/books?hl=en{\&}lr={\&}id={\_}npVDwAAQBAJ{\&}oi=fnd{\&}pg=PP1{\&}dq=milan+cirkovic+great+silence{\&}ots=IevkNh-yO6{\&}sig=FN4gaotKiynll4Q6Fh6lMm-7VMQ}

\bibitem[{{\'{C}}irkovi{\'{c}} {et~al.}(2019){\'{C}}irkovi{\'{c}},
  Vukoti{\'{c}}, \& Stojanovi{\'{c}}}]{Cirkovic2019}
{\'{C}}irkovi{\'{c}}, M.~M., Vukoti{\'{c}}, B., \& Stojanovi{\'{c}}, M. 2019,
  arXiv pre-print, arXiv:1905.03146.
\newblock \url{http://arxiv.org/abs/1905.03146}

\bibitem[{Cocconi \& Morrison(1959)}]{COCCONI1959}
Cocconi, G., \& Morrison, P. 1959, Nature, 184, 844.
\newblock \url{http://www.nature.com/articles/184844a0}

\bibitem[{Daukantas(2017)}]{daukantas2017breakthrough}
Daukantas, P. 2017, Optics and Photonics News, 28, 26

\bibitem[{Davies(2010)}]{Davies2010}
Davies, P. 2010, {The Eerie Silence: Are We Alone in the Universe?} (Penguin
  UK)

\bibitem[{Davies \& Wagner(2013)}]{davies2013searching}
Davies, P., \& Wagner, R. 2013, Acta Astronautica, 89, 261

\bibitem[{Denning(2018)}]{denninghumans}
Denning, K. 2018, in Decoding Alien Intelligence Workshop, SETI Institute

\bibitem[{Drake(1965)}]{Drake1965}
Drake, F. 1965, in Current aspects of exobiology, ed. G.~Mamikunian \& M.~H.
  Briggs (Oxford University Press: Pergamon Press Inc.), 323--345.
\newblock \url{http://adsabs.harvard.edu/abs/1965cae..book..323D}

\bibitem[{Drake(1961)}]{Drake1961}
Drake, F.~D. 1961, Physics Today, 14, 40.
\newblock \url{http://physicstoday.scitation.org/doi/10.1063/1.3057500}

\bibitem[{Drake \& Sagan(1973)}]{drake1973interstellar}
Drake, F.~D., \& Sagan, C. 1973, Nature, 245, 257

\bibitem[{Dyson(1960)}]{Dyson1960}
Dyson, F.~J. 1960, Science, 131, 1667.
\newblock \url{http://science.sciencemag.org/content/131/3414/1667.short}

\bibitem[{Dyson(1966)}]{Dyson1966}
---. 1966, Perspectives in Modern Physics, 641

\bibitem[{Enriquez {et~al.}(2017)Enriquez, Siemion, Foster, Gajjar, Hellbourg,
  Hickish, Isaacson, Price, Croft, DeBoer, {et~al.}}]{enriquez2017breakthrough}
Enriquez, J.~E., Siemion, A., Foster, G., {et~al.} 2017, The Astrophysical
  Journal, 849, 104

\bibitem[{Forgan \& Nichol(2011)}]{Forgan2011}
Forgan, D., \& Nichol, R. 2011, International Journal of Astrobiology, 10, 77.
\newblock
  \url{https://www.cambridge.org/core/product/identifier/S1473550410000236/type/journal{\_}article}

\bibitem[{Frank {et~al.}(2017)Frank, Kleidon, \& Alberti}]{Frank2017}
Frank, A., Kleidon, A., \& Alberti, M. 2017, {Earth as a Hybrid Planet: The
  Anthropocene in an Evolutionary Astrobiological Context},  Elsevier,
  doi:10.1016/j.ancene.2017.08.002.
\newblock
  \url{https://www.sciencedirect.com/science/article/pii/S2213305417300425}

\bibitem[{Freitas(1980)}]{freitas1980interstellar}
Freitas, R. 1980, British Interplanetary Society, Journal(Interstellar
  Studies), 33, 95

\bibitem[{Haqq-Misra \& Kopparapu(2012)}]{Haqq-Misra2012}
Haqq-Misra, J., \& Kopparapu, R.~K. 2012, Acta Astronautica, 72, 15.
\newblock \url{http://arxiv.org/abs/1111.1212
  http://dx.doi.org/10.1016/j.actaastro.2011.10.010}

\bibitem[{Harris(1986)}]{Harris1986}
Harris, M.~J. 1986, Astrophysics and Space Science, 123, 297.
\newblock \url{http://link.springer.com/10.1007/BF00653949}

\bibitem[{Hart(1975)}]{Hart1975}
Hart, M.~H. 1975, Quarterly Journal of the Royal Astronomical Society, 16, 128.
\newblock
  \url{http://articles.adsabs.harvard.edu/full/1975QJRAS..16..128H/0000128.000.html}

\bibitem[{Hippke(2017)}]{Hippke2017}
Hippke, M. 2017, International Journal of Astrobiology, 18, 267.
\newblock
  \url{https://www.cambridge.org/core/product/identifier/S1473550417000507/type/journal{\_}article
  http://arxiv.org/abs/1706.03795}

\bibitem[{Jugaku \& Nishimura(1997)}]{Jugaku1997}
Jugaku, J., \& Nishimura, S. 1997, International Astronomical Union Colloquium,
  161, 707.
\newblock
  \url{https://link.springer.com/content/pdf/10.1007/3-540-54752-5{\_}235.pdf}

\bibitem[{{Kardashev}(1964)}]{kardashev1964}
{Kardashev}, N.~S. 1964, \sovast, 8, 217

\bibitem[{Klein \& Gulkis(1991)}]{Klein1991}
Klein, M.~J., \& Gulkis, S. 1991, in Bioastronomy The Search for
  Extraterrestial Life — The Exploration Broadens (Berlin, Heidelberg:
  Springer Berlin Heidelberg), 203--209.
\newblock \url{http://link.springer.com/10.1007/3-540-54752-5{\_}216}

\bibitem[{Kuiper \& Morris(1977)}]{Kuiper1977}
Kuiper, T.~B., \& Morris, M. 1977, Science, 196, 616

\bibitem[{Lingam \& Loeb(2019)}]{Lingam2018}
Lingam, M., \& Loeb, A. 2019, Astrobiology, 19, 28.
\newblock
  \url{http://arxiv.org/abs/1807.08879{\%}0Ahttp://dx.doi.org/10.1089/ast.2018.1936}

\bibitem[{Lorimer {et~al.}(2007)Lorimer, Bailes, McLaughlin, Narkevic, \&
  Crawford}]{Lorimer2007}
Lorimer, D.~R., Bailes, M., McLaughlin, M.~A., Narkevic, D.~J., \& Crawford, F.
  2007, Science, 318, 777.
\newblock \url{http://www.ncbi.nlm.nih.gov/pubmed/17901298}

\bibitem[{Macmahon {et~al.}(2018)Macmahon, Price, Lebofsky, Siemion, Croft,
  DeBoer, Enriquez, Gajjar, Hellbourg, Isaacson, Werthimer, Abdurashidova,
  Bloss, Brandt, Creager, Ford, Lynch, Maddalena, McCullough, Ray, Whitehead,
  \& Woody}]{Macmahon2018}
Macmahon, D.~H., Price, D.~C., Lebofsky, M., {et~al.} 2018, Publications of the
  Astronomical Society of the Pacific, 130, arXiv:arXiv:1707.06024v2

\bibitem[{McGee(2010)}]{McGee2010}
McGee, B.~W. 2010, Space Policy, 26, 209.
\newblock
  \url{https://www.sciencedirect.com/science/article/pii/S0265964610000858}

\bibitem[{Meadows {et~al.}(2018)Meadows, Reinhard, Arney, Parenteau,
  Schwieterman, Domagal-Goldman, Lincowski, Stapelfeldt, Rauer, DasSarma,
  Hegde, Narita, Deitrick, Lustig-Yaeger, Lyons, Siegler, \&
  Grenfell}]{Meadows2018}
Meadows, V.~S., Reinhard, C.~T., Arney, G.~N., {et~al.} 2018, Astrobiology, 18,
  630.
\newblock \url{http://www.liebertpub.com/doi/10.1089/ast.2017.1727}

\bibitem[{Mix(2019)}]{Mix2019AbSciCon}
Mix, L. 2019, abSciCon Seattle

\bibitem[{{NASA Technosignature Workshop
  Participants}(2018)}]{participants2018nasa}
{NASA Technosignature Workshop Participants}. 2018, arXiv preprint
  arXiv:1812.08681

\bibitem[{Niven(1970)}]{Niven1970}
Niven, L. 1970, {Ringworld} (Ballantine Books), 342.
\newblock
  \url{https://books.google.com/books?hl=en{\&}lr={\&}id=uv4vqKYsyawC{\&}oi=fnd{\&}pg=PA1{\&}dq=ringworld{\&}ots=UOUcsRbdx5{\&}sig=rutn0NlWwJJsjIW25rgY5OTeh9A{\#}v=onepage{\&}q=ringworld{\&}f=false}

\bibitem[{Schwartz \& Townes(1961)}]{SCHWARTZ1961}
Schwartz, R.~N., \& Townes, C.~H. 1961, Nature, 190, 205.
\newblock \url{http://www.nature.com/articles/190205a0}

\bibitem[{Socas-Navarro(2018)}]{socas2018possible}
Socas-Navarro, H. 2018, The Astrophysical Journal, 855, 110

\bibitem[{Stull(1979)}]{Stull1979}
Stull, M. 1979, Journal of the British Interplanetary Society, 32, 221

\bibitem[{Tough(1998)}]{Tough1998}
Tough, A. 1998, Acta Astronautica, 42, 745.
\newblock
  \url{https://www.sciencedirect.com/science/article/pii/S0094576598000356}

\bibitem[{Wright(2018)}]{wright2018prior}
Wright, J.~T. 2018, International Journal of Astrobiology, 17, 96

\bibitem[{Wright {et~al.}(2014)Wright, Griffith, Sigurdsson, Povich, \&
  Mullan}]{Wright2014}
Wright, J.~T., Griffith, R.~L., Sigurdsson, S., Povich, M.~S., \& Mullan, B.
  2014, Astrophysical Journal, 792, 27.
\newblock
  \url{http://stacks.iop.org/0004-637X/792/i=1/a=27?key=crossref.3a5265f423891a6b8b73c59fdf698755}

\bibitem[{Wright {et~al.}(2018{\natexlab{a}})Wright, Kanodia, \&
  Lubar}]{Wright2018a}
Wright, J.~T., Kanodia, S., \& Lubar, E. 2018{\natexlab{a}}, The Astronomical
  Journal, 156, 260.
\newblock
  \url{https://iopscience.iop.org/article/10.3847/1538-3881/aae099/meta}

\bibitem[{Wright {et~al.}(2018{\natexlab{b}})Wright, Sheikh, Alm{\'{a}}r,
  Denning, Dick, \& Tarter}]{Wright2018}
Wright, J.~T., Sheikh, S., Alm{\'{a}}r, I., {et~al.} 2018{\natexlab{b}}, arXiv
  preprint, arXiv:1809.06857.
\newblock \url{http://arxiv.org/abs/1809.06857}

\bibitem[{Wright {et~al.}(2018{\natexlab{c}})Wright, Horowitz, Maire,
  Werthimer, Antonio, Aronson, Chaim-Weismann, Cosens, Drake, Howard,
  {et~al.}}]{wright2018panoramic}
Wright, S.~A., Horowitz, P., Maire, J., {et~al.} 2018{\natexlab{c}}, in
  Ground-based and Airborne Instrumentation for Astronomy VII, Vol. 10702,
  International Society for Optics and Photonics, 107025I

\end{thebibliography}

\end{document}